\documentstyle[12pt,aps,preprint]{revtex}

\preprint{\begin{tabular}{r}
  {\bf hep-ph/0004244}\\
  CERN-TH-2000-122 \\YUMS 00-04
\end{tabular}}

\def\barr{\begin{eqnarray}}
\def\earr{\end{eqnarray}}
\def\beq{\begin{equation}}
\def\eeq{\end{equation}}

\begin{document}

\title
{The CKM phase $\alpha$ through $B \to a_0 \pi$}

\author
{Amol S. Dighe\footnote{amol.dighe@cern.ch}}
\address
{Theory Division, CERN, CH-1211 Geneva 23, Switzerland.}

\author
{C. S. Kim\footnote{kim@kimcs.yonsei.ac.kr,
     ~~http://phya.yonsei.ac.kr/\~{}cskim/}}
\address{Department of Physics and IPAP, Yonsei University, 
     Seoul 120-749,  Korea. \\
Dept of Physics, University of Wisconsin, 1150 University Ave, Madison, WI
53706,  U.S.A.}

\maketitle

\begin{abstract}

We propose the decay modes $B \to a_0 (\to \eta \pi) \pi$ 
to determine the  CKM phase $\alpha$.
One can analyze these modes through
(i) the $B \to a_0 \pi$ isospin pentagon, 
(ii) the time dependent Dalitz plot of 
$B^0(t)  \to a_0^{\pm} \pi^{\mp} \to \eta \pi^+ \pi^-$,
and (iii) the time dependence of 
$B^0(t) \to a_0^0 (\to \eta \pi^0) \pi^0$.
We show that the $a_0 \pi$ modes have 
certain advantages as compared to
the $\rho \pi$ modes, 
and strongly recommend the time dependent
Dalitz plot analysis in the $a_0 \pi$ channel.

\end{abstract}

\section{Introduction}

The determination of the CKM phase $\alpha$ through 
$B \to \rho \pi$ decay modes has been widely discussed. 
The isospin pentagon analysis for the $\rho \pi$ case 
\cite{quinn1} can, in principle, solve for $\alpha$ as
long as the intermediate $\rho$ state can be cleanly
identified and there are no interference effects between
different channels. The interference effects can be 
taken care of by a time dependent Dalitz plot analysis
\cite{quinn2}.
Some of the $\rho \pi$ modes have already
been observed at CLEO \cite{cleo}, and
the experimental feasibility studies of these modes 
have been performed for the $B$ factories 
\cite{babar} and the hadronic machines \cite{cdf,lhc}.
The theoretical 
issues involving the measurement of the phase of the
$t$-quark penguin through these modes have also 
been addressed \cite{rahul}.

Although 
$B \to \rho \pi$ are the most favored modes for
determining $\alpha$, their utility is hampered by a few 
factors. 
Since all the final state particles are pions and 
the width of $\rho$ is large ($\approx 150$ MeV), the combinatorial
background prevents one from a clean separation between
the $\rho^+ \pi^-$, $\rho^- \pi^+$ and $\rho^0 \pi^0$
channels. The isospin pentagon analysis, as proposed in \cite{quinn1},
then cannot be carried out and one has to use the full
three body analysis \cite{quinn2} taking into account the 
interference effects between different channels.
Neglecting the non-resonant contributions and 
electroweak penguins, this analysis can be performed
in principle to give $\alpha$ without discrete ambiguities. 
However, this involves performing a likelihood fit with
as many as nine independent parameters, and the presence of
resonances near $\rho$ which decay to $\pi \pi$ also
restricts the potential of this mode.

In this article, we point out that analyses of the decay modes 
$B \to a_0 \pi$  offer certain advantages over 
the $\rho \pi$ modes. The combinatorial background and
the background due to nearby resonances is smaller, and the 
Dalitz plot analysis requires a likelihood fit with 
only seven independent parameters, as compared to nine
in the case of the $\rho \pi$ channel.
If the branching fractions of $B$ into these two channels
are similar (which they are expected to be), 
the $a_0 \pi$ modes should be
able to perform as well as $\rho \pi$ modes,
and possibly even better, for the determination of $\alpha$.

\section{The isospin ``pentagon'' analysis}

We shall use the notation anologous to the one
introduced in \cite{quinn1} for the amplitudes:
the amplitudes for the five $B \to a_0 \pi$
processes are

\barr
\sqrt{2} A(B^+ \to a_0^+ \pi^0) & = S_1 = &
T^{+0} + 2 P_1 ~~, \nonumber \\
\sqrt{2} A(B^+ \to a_0^0 \pi^+) & = S_2 = &
T^{0+} - 2 P_1  ~~,\nonumber \\
A(B^0 \to a_0^+ \pi^-) & = S_3 = &
T^{+-} +  P_1 + P_0  ~~,\nonumber \\
A(B^0 \to a_0^- \pi^+) & = S_4 = &
T^{-+} -  P_1 +P_0  ~~,\nonumber \\
2 A(B^0 \to a_0^0 \pi^0) & = S_5 = &
T^{+0} + T^{0+} - T^{+-} - T^{-+} -2 P_0  ~~.
\earr
For the $CP$-conjugate processes, we define the amplitudes
$\overline{S}_i$, $\overline{T}^{ij}$ and $\overline{P}_i$ which differ
from the original amplitudes only in the sign of the 
weak phase of each term.

The isospin pentagon analysis 
with the decays $B \to a_0 \pi$ 
can be carried out as outlined in \cite{quinn1}:
the decay rates of all the five modes
above (and their $CP$-conjugate modes) can be measured.
Adding to this the time-dependent $CP$ asymmetries in the
decays of neutral $B$ mesons
($\stackrel{(-)}{B^0}(t) \to a_0^+ \pi^-, ~a_0^- \pi^+, ~a_0^0 \pi^0$),
we have 12 observables
(taking into account an arbitrary normalization),
which in principle can solve for the 12 unknowns:
four $T^{ij}$, two $P_i$, five relative strong phases
between them, and the CKM phase $\alpha$. 

The isospin pentagon analysis for the $B \to a_0 \pi$ mode
can be carried out more cleanly than that in the
$B \to \rho \pi$ mode, because of the following reasons:
\begin{itemize}

\item The combinatorial background is smaller:
the final state
in $a_0 \pi \to \eta (\to \gamma \gamma) \pi \pi$ 
consists of two photons and two pions, and given
the small width of $a_0$, the probability of both the
$\eta \pi$ pairs having an invariant mass below
the $a_0$ peak is small. For the same reason,
the contamination due to the 
$\eta \rho$ resonance will be small since this
resonance band ``intersects'' the $a_0 \pi$ resonance
band in the Dalitz plot, and does not ``overlap'' it.

\item The background due to nearby resonances is smaller:
the nearby resonances that decay to $\pi \pi$ 
({\it e.g.} $f_0, \omega, \sigma$) can contribute to the events 
below the $\rho$ peak. On the other hand, 
the nearest resonance to $a_0$(980) that can decay to $\eta \pi$ 
is $a_2(1320)$, which also has a small width ($\approx 110$ MeV)
and has a branching ratio of about 15\%, so that its
contribution to $\eta \pi$ below the peak of $a_0$
is small. 

\end{itemize}

The discrete ambiguities in the determination of $\alpha$ 
through the pentagon analysis \cite{gronau}
cannot be avoided, and the interference effects
might restrict the efficiency of this method. In that case,
the full three body analysis described below has to be employed.

\section{The time dependent three body Dalitz plot analysis}

The Dalitz plot analysis can take care of the interference
terms between two decay channels leading to the same final
state. 
This analysis is more general than the
isospin pentagon analysis and can be used even in the case of
small interference effects, though substantial interference 
effects increase the efficacy of this analysis.

The analysis, similar to the one 
suggested in \cite{quinn2}, can be carried out in the 
following manner: ignoring the non-resonant contributions
and denoting
\beq
f^i \equiv A(a^i_0 \to \eta \pi^i)~~,
\eeq
we can write
\beq
A(B^0 \to \eta \pi^+ \pi^-) =  f^+ S_3 + f^- S_4 ~~,
\eeq
and its $CP$-conjugate decay amplitude
\beq
A(\overline{B}^0 \to \eta \pi^+ \pi^-)  =  
f^- \overline{S}_3 + f^+ \overline{S}_4 ~~.
\eeq
Using the time dependent decay amplitude
\barr
A(B^0(t) \to \eta \pi^+ \pi^-)  & = & 
e^{-\Gamma t /2} \times \nonumber \\ 
 & \times & \left[ \cos(\frac{\Delta m t}{2}) A(B^0 \to \eta \pi^+ \pi^-) 
+ i q \sin(\frac{\Delta m t}{2}) A(\overline{B}^0 \to \eta \pi^+ \pi^-) \right]~~,
\earr
where $q$ is defined such that the mass eigenstates of neutral $B$ are
\beq
B_H = (B^0 + q \overline{B})/\sqrt{2} ~~, ~~
B_L = (B^0 - q \overline{B})/\sqrt{2}~~,
\eeq
we can write the time dependent decay rate as
\beq
\Gamma(B^0(t) \to \eta \pi^+ \pi^-) =
e^{-\Gamma t} \sum_k {\cal F}_k  \left[
{\cal A}_{k1} + {\cal A}_{kC} \cos (\Delta m t) +
{\cal A}_{kS} \sin (\Delta m t)  \right]~~.
\label{fadist}
\eeq
The expressions for ${\cal F}_{k}$ and ${\cal A}_{ki}$
(where $i \in \{ 1,C,S \}$) are given in Table~\ref{fatable}.

The strong decay amplitudes $f^j$ s ($j \in \{ +,-,0 \}$)
are functions of the
invariant mass $m_j$ of the $\eta \pi^j$ pair.
They can be naively approximated by the Breit - Wigner 
function as
\beq
f^j(m_j) = \frac{\Gamma_{a_0}}{2(m_{a_0} - m_j) - i \Gamma_{a_0}}~~.
\eeq 
However, the $K \bar{K}$ threshold near $a_0$(980) 
distorts the spectrum of $f^j(m_j)$ \cite{pdg}.
To take care of this, one has to either 
(i) select only those events below a certain value of $m_j$
(such that the line shape of the $\eta \pi^j$ resonance
in this region is not affected much 
by the $K \bar{K}$ threshold effects) or
(ii) use a coupled channel model to theoretically determine
the line shape \cite{coupled}.

Using the time evolution (\ref{fadist}), 
the coefficients of ${\cal F}_k$ can be separated into 
${\cal A}_{k1}$, ${\cal A}_{kC}$ and ${\cal A}_{kS}$,
which are terms bilinear in the amplitudes $S$ and $\overline{S}$.
The $\alpha$ dependence of the leading order 
terms in ${\cal A}_{ki}$ 's
has been shown in Table~\ref{fatable}. It shows that both
$\sin(2 \alpha)$ and $\cos(2\alpha)$ can be measured 
independently, and hence the discrete ambiguity 
($\alpha \to \pi/2 - \alpha$) is absent (this has been
noticed in \cite{quinn2} for the $\rho \pi$ mode).

The decay $B^0 \to \eta \pi^+ \pi^-$ involves the contribution from
only two channels: $a_0^+ \pi^-$ and $a_0^- \pi^+$ (as
opposed to the $B^0 \to \pi^+ \pi^- \pi^0$ decay, which
gets the contribution from three channels:
$\rho^+ \pi^-$, $\rho^0 \pi^0$ and $\rho^- \pi^+$).
Therefore, the number of parameters involved in the 
$a_0 \pi$ analysis is smaller (seven as compared to nine for 
$\rho \pi$, if we take
the values of the masses and widths of $\rho$ and $a_0$ to be
known). Moreover, the smaller width of $a_0$ implies
a significant interference-free region in the Dalitz plot, 
which would allow the determination of the coefficients of
$f^+ f^{+*}$ and $f^- f^{-*}$ (See Table \ref{fatable})
independently. This makes the likelihood fit more robust.

We also have the corresponding amplitudes for
decays to all neutral particles:
\barr
A(B^0 \to \eta \pi^0 \pi^0) & = & f^0 S_5/2 ~~, \nonumber \\
A(\overline{B}^0 \to \eta \pi^0 \pi^0) & = & f^0 \overline{S}_5/2 ~~.
\earr
The time dependent decay rate is
\barr
\Gamma(B^0(t) &\to& \eta \pi^0 \pi^0) =  e^{-\Gamma t} 
~\frac{f^0 f^{0*}}{4} \nonumber \\
 & \times & \left[ \frac{|S_5|^2 + |\overline{S}_5|^2}{2} + 
\frac{|S_5|^2 - |\overline{S}_5|^2}{2} \cos(\Delta m t) -
{\rm Im} (q \overline{S}_5 S_5^*) \sin(\Delta m t) \right].
\label{pi0-s2alpha}
\earr
The term ${\rm Im} (q \overline{S}_5 S_5^*)$ depends on 
$\sin(2\alpha)$ in the leading order. The three terms
above are sufficient to determine the value of 
$\sin(2\alpha)$ in the approximation of small 
QCD penguins, but the expected smallness of the branching
fraction [${\cal O}(10^{-6})$] 
would mean insufficient statistics for the
determination of $\alpha$.
If the penguin contribution is substantial
or the assumption of colour suppression is not valid,
the mode $B \to \eta \pi^0 \pi^0$ may
still have a larger branching fraction than 
naively expected, though.

In the case of the charged $B$ decays, we have
\beq
A(B^+ \to \eta \pi^+ \pi^0) =  f^+ S_1 + f^0 S_2 ~~,
\eeq
and its $CP$-conjugate decay amplitude
\beq
A(B^- \to \eta \pi^- \pi^0)  =  
f^- \overline{S}_1 + f^0 \overline{S}_2 ~~.
\eeq
It is possible to observe the direct $CP$ violation, but its
measurement cannot be connected to $\alpha$ without the 
knowledge of the strong phases.

\section{Discussions and conclusions}

The good pion reconstruction in the $B$ factories
makes the study of multi-pion modes at these experiments 
attractive ({\it e.g.}  $4\pi$ modes like $\rho \rho, a_1 \pi$). 
But the problems of combinatorial background in these
modes become even more severe than $\rho \pi$. In addition,
since one has to deal with a four particle phase space, the
overlap regions of the resonance bands are smaller than
those of $\rho \pi$, which reduces the impact of the
interference effects \cite{babar}.
In the case of $a_0 \pi$ one has to deal with only a
three particle phase space like in $\rho \pi$, since the 
two photons in the final state have to come from $\eta$.
So it is free of the phase space problems with the
$4\pi$ modes.

The $B$ factory experiments also have an environment 
clean enough to
reconstruct photons and consequently $\eta$, so the detection of
$B \to a_0 \pi$ should not pose a hard challenge.
One loses a bit on the branching ratio 
($B(\eta \to \gamma \gamma) \approx 39$\%), but that is partially
compensated for by
the fact that the photons from $\eta$ are on an average 
more well separated than those from $\pi^0$, which would also imply less
background for $\eta$.
The branching ratios ${\cal B}(B \to a_0^i \pi^j)$ 
are expected to be nearly the same as the branching ratios 
${\cal B}(B \to \rho^i \pi^j)$, {\it i.e.} ${\cal O}(10^{-5})$
when at least one of the final state pions is charged. 
The Dalitz plot analysis of $B \to \eta \pi^+ \pi^-$ needs
only charged pions, so the statistics available will be
comparable to that of the $\rho \pi$ mode.

The $a_0^{\pm} (\to \eta \pi^\pm) \pi^\mp$ modes can also be
combined with the $\rho^0(\to \pi^+ \pi^-) \eta$ mode
on the same Dalitz plot, and the additional interference
regions can be used to obtain supplementary information
on $\alpha$ and the magnitudes of the tree and penguin
amplitudes \cite{durham}.

As a final note, we give one short comment on the possible extension to the 
$B \to \eta \pi \pi$ isospin triangle analysis:
since the isospin of $\eta$ is zero, in the leading order
the isospin analysis
of $B \to \eta \pi \pi$ can be the same as the isospin analysis of
$B \to \pi \pi$, and the CKM angle $\alpha$ can be determined
from the amplitude triangle
\beq
\frac{1}{\sqrt{2}} A(B^0 \to \eta \pi^+ \pi^-) +
A(B^0 \to \eta \pi^0 \pi^0) = 
A(B^+ \to \eta \pi^+ \pi^0) 
\eeq
and its $CP$-conjugate triangle
\beq
\frac{1}{\sqrt{2}} A(\overline{B}^0 \to \eta \pi^+ \pi^-) +
A(\overline{B}^0 \to \eta \pi^0 \pi^0) = 
A(B^- \to \eta \pi^- \pi^0) 
\eeq
in the same manner as has been proposed for $\pi \pi$ in \cite{gl}.
However, since unlike in the $B \to \pi \pi$ modes,
the isospin of $\pi-\pi$ can be $I=1$ 
in the $B \to \eta \pi \pi$ modes (when the $\pi-\pi$ state has $L=1$ 
and is in a relative $P$-wave with $\eta$),  
although such final states are suppressed because of their higher
angular momenta.
If the analysis at the $B$ factories can disentangle such final states
(through angular distributions, for example), then even the 
$B \to \eta \pi \pi$ isospin triangle analysis may be performed.

In conclusion, we have shown that the modes 
$B \to a_0 (\to \eta \pi) \pi$ 
can be used to determine the value of $\alpha$ through
(i) the isospin pentagon analysis of $B \to a_0 \pi$,
(ii) the time dependent Dalitz plot analysis of
$B^0(t) \to \eta \pi^+ \pi^-$,
and (iii)  the time dependence of the decay
$B^0(t) \to \eta \pi^0 \pi^0$. 
The $a_0 \pi$ modes have less background
than the $\rho \pi$ modes, and the Dalitz plot analysis for
$B^0(t) \to a_0^{\pm} \pi^{\mp} \to \eta \pi^+ \pi^-$
involves two less parameters than the 
corresponding $\rho \pi$ analysis. The 
larger interference-free region in the $a_0 \pi$ case
would also make the likelihood fit more robust.
Since the branching ratios for these two sets of modes are
expected to be similar, the $a_0 \pi$ modes should be
able to perform at least as well as the $\rho \pi$ modes.
We, therefore, strongly propose that the 
$B \to a_0 (\to \eta \pi) \pi$
modes be studied at the $B$ factories and the 
upcoming hadron colliders.

\newpage

\section*{Acknowledgements}

We are grateful to G. Buchalla and H. Quinn for going through
this manuscript and offering valuable comments. 
We would also like to thank C. Goebel, P. Galumian, 
M. Smizanska and H. Yamamoto for useful discussions.
The work of C.S.K. was supported 
by the KRF Grants (Project No. 1997-011-D00015 and
Project No. 2000-015-DP0077).

\newpage

\begin{table}
\begin{center}
\begin{tabular}{|cccc|}
%\hline
${\cal F}_{k}$ & $i$ & ${\cal A}_{ki}$ & $\alpha$ dependence \\
\hline
$f^+ f^{+*}$ & $1$ & $(S_3 S_3^* + \overline{S}_4 \overline{S}_4^*)/2$ & $1$ \\
& $C$ & $(S_3 S_3^* - \overline{S}_4 \overline{S}_4^*)/2$ & $1$ \\
& $S$ & $-{\rm Im}(q \overline{S}_4 S_3^*)$ & $\sin(2\alpha)$ \\
\hline
$f^- f^{-*}$ & $1$ & $(S_4 S_4^* + \overline{S}_3 \overline{S}_3^*)/2$ & $1$ \\
& $C$ & $(S_4 S_4^* - \overline{S}_3 \overline{S}_3^*)/2$ & $1$ \\
& $S$ & $-{\rm Im}(q \overline{S}_3 S_4^*)$ & $\sin(2\alpha)$ \\
\hline
${\rm Re}(f^+ f^{-*})$ & $1$ & 
${\rm Re}(S_3 S_4^* + \overline{S}_4 \overline{S}_3^*)$ & $1$ \\
& $C$ & ${\rm Re}(S_3 S_4^* - \overline{S}_4 \overline{S}_3^*)$ & $1$ \\
& $S$ & $-{\rm Im}(q \overline{S}_4 S_4^* - q^* S_3 \overline{S}_3^*)$ &
$\sin(2\alpha)$ \\
\hline
${\rm Im}(f^+ f^{-*})$ & $1$ & 
$- {\rm Im}(S_3 S_4^* + \overline{S}_4 \overline{S}_3^*)$ & $1$ \\
& $C$ & $-{\rm Im}(S_3 S_4^* - \overline{S}_4 \overline{S}_3^*)$ & $1$ \\
& $S$ & $-{\rm Re}(q \overline{S}_4 S_4^* - q^* S_3 \overline{S}_3^*)$ &
$\cos(2\alpha)$ \\
%\hline
\end{tabular}
\end{center}
\caption{ The values of ${\cal F}_{k}$ and ${\cal A}_{ki}$
in the distribution of events (see eq.~\ref{fadist})
in $B^0 \to \eta \pi^+ \pi^-$. The last column shows
the $\alpha$ dependence of the leading order term in
${\cal A}_{ki}$, {\it i.e.} when all $P_i$ are set to zero.
\label{fatable}}
\end{table}

\end{document}